\documentclass[a4paper,12pt]{article}
\usepackage[utf8]{inputenc}
\usepackage{amsmath}
\usepackage{amsfonts}
\usepackage{amssymb}
\usepackage{amsthm}
\usepackage{caption}
\usepackage{multirow}
\usepackage{fontenc}
\usepackage{graphicx}
\usepackage{color}
\usepackage[breaklinks,colorlinks,allcolors=blue,linkcolor=red]{hyperref}
\usepackage{amsmath,amsfonts,amssymb,amsmath,latexsym,mathrsfs}

\usepackage[symbol]{footmisc}
\usepackage[total={6in, 9.5in}]{geometry}

\date{}

\def\title#1{\begin{center} {\LARGE #1 \vspace{0.5cm}} \end{center}}
\def\author#1{\begin{center}{ \large #1} \end{center}}
\def\affil#1{\begin{center}{ \it #1} \end{center}}
\def\email#1{\begin{center}{\normalsize #1} \end{center}}
\def\preprint#1{\rightline{\begin{tabular}{l} #1 \end{tabular}}}
\def\acknowledgement#1{{\noindent \bf\large Acknowledgements.} \,\,#1 }

\begin{document}
\preprint{KCL-2020-12} % you may put your preprint number here, otherwise leave it blank

\title{Could the Hubble Tension be Pointing Towards the Neutrino Mass Mechanism?} %put your title here

\author{Miguel Escudero Abenza$^{1\,\dagger}$\footnote[5]{Speaker}\,\,, Samuel J. Witte$^{2\,\ast}$} %you may use your own symbol for representing email

\affil{$^1$Department of Physics, King's College London, United Kingdom   \\ 
$^2$IFIC \& Universidad de Valencia, CSIC-UVEG, Spain \\}

\email{$^{\dagger}$miguel.escudero@kcl.ac.uk,\, $^{\ast}$sam.witte@ific.uv.es }
\vspace{1cm}
%
%
%%%%%%%%%%%%%%%%%%%%%%%%%%%%%%%%%%%%% Abstract begins here %%%%%%%%%%%%%%%%%%%%%%%%%%%%%%%%%%%%%%%%%%%%%%%%%%%%%%%%%%%%%%%%%%%%%%%%%%%%%%%%
%%%%%%%%%%%%%%%%%%%%%%%%%%%%%%%%%%%%%%%%%%%%%%%%%%%%%%%%%%%%%%%%%%%%%%%%%%%%%%%%%%%%%%%%%%%%%%%%%%%%%%%%%%
\begin{abstract}
Local measurements of the Hubble constant currently disagree with the high-precision value that is inferred from the CMB under the assumption of a $\Lambda$CDM cosmology. The significance of this tension clearly motivates studying extensions of the standard cosmological model capable of addressing this outstanding issue. Broadly speaking, models that have been successful in reducing the the tension between the CMB and local measurements (without introducing additional tension in other datasets) require an additional component of the energy density in the Universe at a time close to recombination.

In this contribution, I will show that the Majoron -- a pseudo-Goldstone boson arising from the spontaneous breaking of a global lepton number symmetry and often associated with the neutrino mass mechanism -- can help to reduce the Hubble tension. Importantly, I will also show that current CMB observations can constrain neutrino-Majoron couplings as small as $10^{-13}$, which within the type-I seesaw mechanism correspond to scales of lepton number breaking as high as $\sim 1\,\text{TeV}$. 
\end{abstract}

%%%%%%%%%%%%%%%%%%%%%%%%%%%%%%%%%%%%%%%%%%%%
\vspace*{4cm}
\begin{center}
{\LARGE{Poster and Talk Presented at}}\\
\vspace{1cm}
\Large{NuPhys2019: Prospects in Neutrino Physics\\
Cavendish Conference Centre, London, 16--18 December 2019}
\end{center}
%%%%%%%%%%%%%%%%%%%%%%%%%%%%%%%%%%%%%%%%%%%%

%%%%%%%%%%%%%%%%%%%%%%%%%%%%%%%%%%%%%%%%%%%%
\begin{center}
\large{\textit{Runner-up Theoretical Poster}, check it out following this~\href{https://www.dropbox.com/s/n6jizypa0wfcfto/Escudero_Poster_Nuphys2019.pdf?dl=0}{LINK}}
\end{center} 
 %%%%%%%%%%%%%%%%%%%%%%%%%%%%%%%%%%%%%%%%%%%%
 
\clearpage

%%%%%%%%%%%%%%%%%%%%%%%%%%%%%%%%%%%%%%%%%%%%
\section{Introduction: The Hubble Tension}
%%%%%%%%%%%%%%%%%%%%%%%%%%%%%%%%%%%%%%%%%%%%
\vspace{-0.1cm}
Local determinations of the expansion rate of the Universe seem to indicate that the Universe is expanding faster than would be expected within the standard cosmological model, $\Lambda$CDM. The rate of expansion of the Universe today is parametrized by Hubble's constant, $H_0$. At present, many local measurements of the Hubble constant\footnote{The only, but notable, exception is the local $H_0$ measurement from the Tip of the Red Giant Branch~\cite{Freedman:2019jwv}. This measurement is only discrepant at the $1.2\sigma$ level with the $H_0$ value inferred from Planck CMB observations within $\Lambda$CDM.} yield values of $H_0$ that are $\sim 4\!-\!6 \sigma$ higher than the value that can be inferred from early Universe probes within the framework of $\Lambda$CDM. This inconsistency goes by the name of the `Hubble tension', see~\cite{Verde:2019ivm} for a more or less up-to-date review. While this problem persists through a variety of different datasets, the largest tension is between the value of $H_0$ obtained from using cepheids to calibrate the distance to type-Ia supernova: $H_0 = 74.0  \pm 1.4\,\text{km}/\text{s}/\text{Mpc}$~\cite{Riess:2019cxk}, with the value inferred from very precise Cosmic Microwave Background (CMB) observations by the Planck satellite within the framework of $\Lambda$CDM: $H_0 = 67.36\pm 0.54\,\text{km}/\text{s}/\text{Mpc}$~\cite{Aghanim:2018eyx}. When these two values are compared at face value, the tension is at $4.4\sigma$. There are two possible avenues to make sense of this tension. First, it is possible that there are unaccounted systematic effects in the local $H_0$ and/or Planck data. However, the data sets and analyses pipelines have gone under an intense scrutiny by the Cosmology community, and so far no one has been able to find any relevant systematic -- this is particularly the case for Planck CMB observations~\cite{Akrami:2018vks}. Second, it is possible that early and late probes of the Universe are unaffected by systematics and are both correct. If this is the case, the tension must be pointing towards physics not accounted for in the $\Lambda$CDM framework. The relevant question is: how should one modify $\Lambda$CDM in order to accommodate the Hubble tension, while simultaneously not spoiling the wide array of observations that are successfully described by the model? It is now widely believed that the most likely modification of $\Lambda$CDM capable of substantially ameliorating/solving the tension is one in which the expansion history of the Universe is enhanced with respect to $\Lambda$CDM prior (and likely close) to recombination~\cite{Knox:2019rjx}. However, simple extensions of the model featuring such an enhanced expansion history -- such as $\Lambda$CDM+$\Delta N_{\rm eff}$ -- are not favored by Planck CMB observations~\cite{Aghanim:2018eyx}.

 Clearly, if the Hubble tension is confirmed, it will require new physics beyond $\Lambda$CDM. Given that neutrino oscillations are the only laboratory evidence of physics beyond the Standard Model it is natural to ask if the Hubble tension can be connected with the neutrino mass mechanism. In~\cite{Escudero:2019gvw}, Sam Witte and myself explored this possibility. In particular, we considered the cosmological implications of a scenario in which lepton number is spontaneously broken, subsequently generating neutrino masses via the type-I seesaw mechanism. In what follows, I will describe the setup considered in~\cite{Escudero:2019gvw}, and summarize the main results obtained in that reference.

\vspace{-0.3cm}

%%%%%%%%%%%%%%%%%%%%%%%%%%%%%%%%%%%%%%%%%%%%
\section{The Majoron: $m_\nu$ within a global $U(1)_L$ }
%%%%%%%%%%%%%%%%%%%%%%%%%%%%%%%%%%%%%%%%%%%%
\vspace{-0.1cm}
Neutrinos are massless in the Standard Model. Perhaps, the most simple and elegant way of generating neutrino masses is via the type-I seesaw mechanism. The key ingredient of the type-I seesaw mechanism is the addition of electroweak-singlet (right handed) neutrinos with a Majorana mass that explicitly breaks lepton number, $M_N$. Upon electroweak symmetry breaking, active neutrinos naturally obtain a small mass $m_\nu \sim y_N^2 v_H^2/M_N$, where $y_N$ is the Higgs-lepton-sterile neutrino Yukawa coupling. 

Within the type-I seesaw, $M_N$ is just a parameter in the Lagrangian. However, our understanding of mass generation suggests that $M_N$ should arise from the spontaneous breakdown of a lepton number symmetry. This was precisely the set-up considered back in 1981 in Ref.~\cite{Chikashige:1980ui}. In this scenario, $U(1)_L$ is a global symmetry, and therefore upon spontaneous symmetry breaking a \textit{pseudo-Goldstone boson} appears in the spectrum: the Majoron ($\phi$). Neutrino-Majoron interactions are described by
\begin{align} 
\mathcal{L}_{\rm int} =  i\, \frac{\lambda}{2}\, \phi\, \bar{\nu}\, \gamma_5\, \nu\,,
\end{align}
where within the type-I seesaw this interaction effectively arises from $\nu\!-\!N$ mixing and is very feeble: $\lambda = 2\, m_\nu/v_L \simeq 10^{-13} \frac{m_\nu}{0.05\,\text{eV}}\frac{1\,\text{TeV}}{v_L}$. The interaction of Majorons with ordinary matter is even weaker. It is loop and neutrino mass suppressed: $\lambda_{ee\phi} \lesssim 10^{-20}$, making the Majoron an extremely elusive particle. 

Quantum Gravity is expected to break all global symmetries and we therefore expect $m_\phi \neq 0$. However, the exact way in which Gravity breaks global symmetries is highly unclear. One may naively expect the Majoron mass to arise from dimension-5 Planck suppressed operators that explicitly break $U(1)_L$. If that is the case, they point towards $m_\phi^{\rm Dim-5} \sim \sqrt{v_H^3/M_{\rm Pl}} \sim 0.1\,\text{keV}$ (see green region of Figure~\ref{fig:Money}). 

%%%%%%%%%%%%%%%%%%%%%%%%%%%%%%%%%%%%%%%%%%%%
\section{Cosmological Implications and Methodology}
%%%%%%%%%%%%%%%%%%%%%%%%%%%%%%%%%%%%%%%%%%%%

Given the weakness of neutrino-Majoron interactions, the only cosmologically relevant processes in which Majorons participate are $\phi \to \bar{\nu}\nu$ and $\bar{\nu}\nu \to \phi$ decays. Note that bosons interacting with neutrinos with coupling strengths of $\lambda \sim 10^{-13}$ and $m_\phi \sim 0.1\,\text{keV}$ have lifetimes of $\tau_\phi \sim 400\,\text{yr} \, \left({0.1\,\text{keV}}/{m_\phi}\right) \, \left({10^{-13}}/{\lambda}\right)^2 $. These lifetimes clearly suggest that Majorons can have important implications for CMB observations. The cosmological implications of Majorons were first highlighted in Ref.~\cite{Chacko:2003dt} and are:
\begin{enumerate}
\item \textit{Neutrino-Majoron interactions induce an enhanced expansion history}. \\ This arises as a result of the fact that across wide regions of parameter space majorons thermalize with neutrinos while relativistic but then decay back into neutrinos while non-relativistic heating up the neutrino fluid. This leads to an enhanced expansion history since $H \propto \sqrt{\rho}$ which is relevant at $T_\nu \lesssim m_\phi$.   
\item \textit{Neutrino-Majoron interactions reduce neutrino freestreaming}. \\ Neutrinos represent 40\% of the energy density of the Universe between $e^+e^-$ annihilation until almost matter-radiation equality. Therefore, neutrino perturbations have a strong impact on the metric perturbations which are the source of the CMB spectra~\cite{Bashinsky:2003tk}. 
\end{enumerate} 

\noindent In~\cite{Escudero:2019gvw}, we were the first to contrast the cosmological implications of Majorons against CMB observations. For that purpose: we \textit{i)}  modeled the two effects outlined above by solving for the background neutrino-Majoron thermodynamics using the methods developed in~\cite{Escudero:2018mvt,Escudero:2020dfa} (\href{https://github.com/MiguelEA/nudec_BSM}{\texttt{NUDEC\_BSM}}), \textit{ii)} implemented the background evolution and  included $\nu\!-\!\phi$ interactions in the Boltzmann hierarchy of neutrinos in the Boltzmann code \href{https://class-code.net/}{\texttt{CLASS}}~\cite{Blas:2011rf}, and \textit{iii)} performed a full MCMC with \href{https://github.com/brinckmann/montepython_public}{\texttt{MontePython}}~\cite{Audren:2012wb,Brinckmann:2018cvx} and used the latest Planck2018 CMB observations~\cite{Akrami:2018vks} to constrain $\lambda$ and $m_\phi$.

%%%%%%%%%%%%%%%%%%%%%%%%%%%%%%%%%%%%%%%%%%%%
\section{Main Results}
\begin{figure}[t]
\centering
\includegraphics[width=1.0\textwidth]{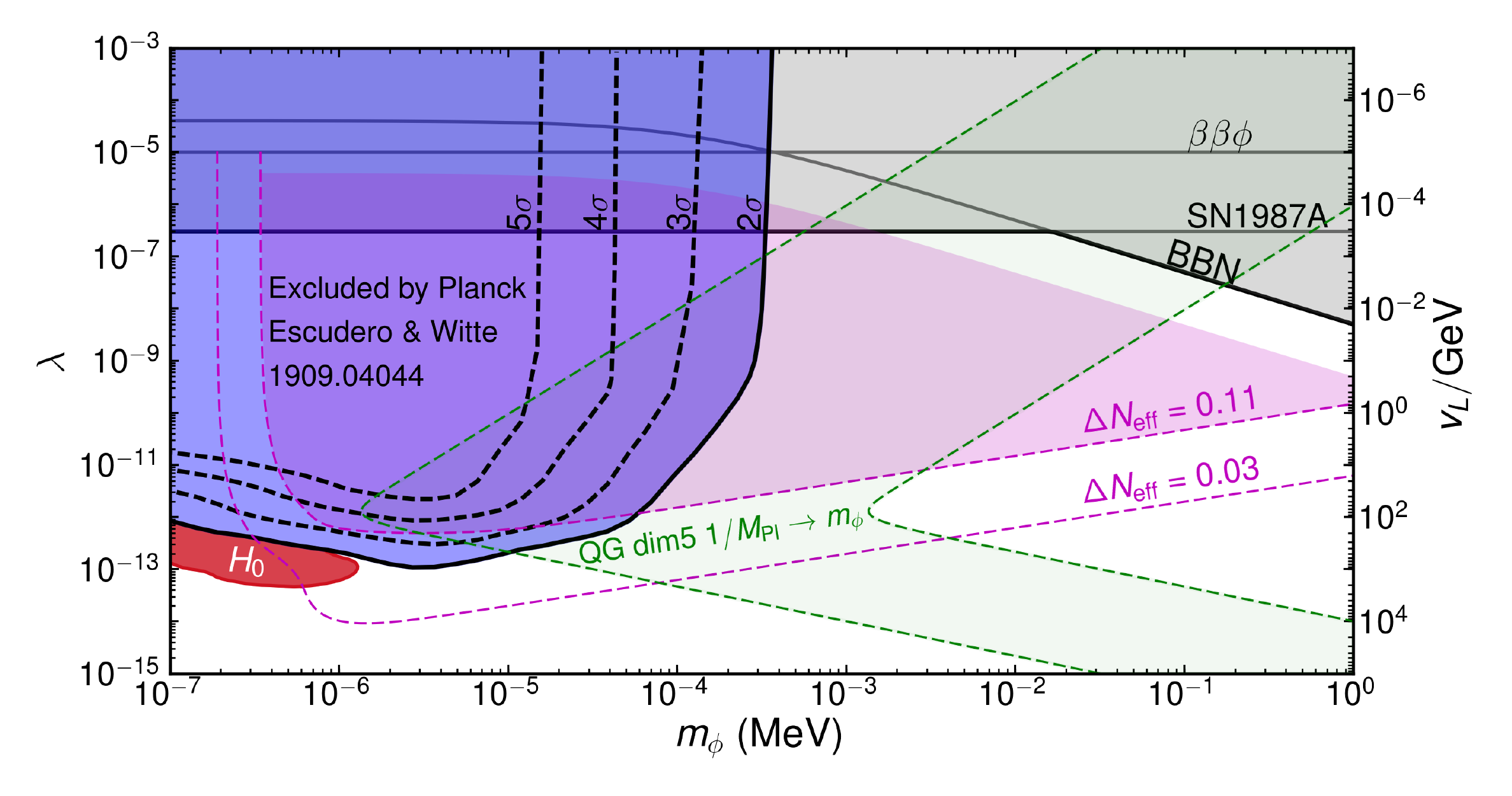}
  \vspace{-1cm}
  \caption{Plot highlighting the parameter space in the Majoron model. Adapted from Figure 1 in~\cite{Escudero:2019gvw} to include $2,\,3,\,4,\,5\sigma$ exclusion limits. Grey contours include bounds from BBN, SN1987A, and $\beta \beta \phi$ decays. The green contour corresponds to the region of parameter space in which the Majoron mass could arise from dimension-5 Planck suppressed operators explicitly breaking lepton number. The blue contour is excluded by Planck legacy CMB observations. In red we highlight the $1\sigma$ preferred region of parameter space for solving the $H_0$ tension within a Majoron+$\Delta N_{\rm eff}$ cosmology.   } \label{fig:Money}
 \end{figure}

%%%%%%%%%%%%%%%%%%%%%%%%%%%%%%%%%%%%%%%%%%%%
\vspace{-0.2cm}
In Figure~\ref{fig:Money} the main results obtained in~\cite{Escudero:2019gvw} are highlighted. Firstly, the pink contour reflects the region of parameter space in which neutrino-majoron interactions render $\Delta N_{\rm eff} \simeq 0.11$ as relevant for CMB observations. Secondly, the blue contour shows the region of parameter space excluded by Planck legacy observations at 95\% CL. We notice that neutrino-majoron coupling strengths as small as $\lambda \sim 10^{-13}$ are probed by current CMB observations. This corresponds to scales of lepton number breaking as high as $v_L \sim 1\,\text{TeV}$ within the type-I seesaw neutrino mass mechanism. Figure~\ref{fig:Money} includes the 2, 3, 4 and 5$\sigma$ exclusion contours to illustrate the constraining power of Planck observations to Majorons in the mass window $0.1 \,\text{eV}\lesssim m_\phi \lesssim 300 \,\text{eV}$. In the region of parameter space highlighted in blue, $\nu\!-\!\phi$ interactions would distort the neutrino perturbations significantly beyond what is allowed by Planck legacy data. 

By performing a full MCMC analysis of the Majoron cosmology we found that the enhanced expansion history within the majoron cosmology is not enough to substantially ameliorate the Hubble tension. We considered the case of a cosmology featuring the Majoron and extra dark radiation parametrized by $\Delta N_{\rm eff}$. For this cosmology, by doing a joint fit to Planck2018+BAO data we found that the Hubble tension can be reduced from $4.4\sigma$ to $2.5\sigma$. The red region of parameter space in Figure~\ref{fig:Money} shows the preferred region of parameter space for solving the Hubble tension within this cosmology. This corresponds, at $1\sigma$, to $\Delta N_{\rm eff} = 0.5\pm0.2$. Very importantly, and unlike in a $\Lambda$CDM+$\Delta N_{\rm eff}$ cosmology, in the Majoron+$\Delta N_{\rm eff}$ case such rather large values of $\Delta N_{\rm eff}$ do not degrade the Planck fit thanks to the presence of majoron-neutrino interactions, see Table I of~\cite{Escudero:2019gvw}.

 %%%%%%%%%%%%%%%%%%%%%%%%%%%%%%%%%%%%%%%%%%%%
\section{Conclusions and Outlook }
%%%%%%%%%%%%%%%%%%%%%%%%%%%%%%%%%%%%%%%%%%%%
The main conclusions obtained in~\cite{Escudero:2019gvw} are:
\begin{enumerate}
\item As highlighted in blue in Figure~\ref{fig:Money}, Planck CMB data strongly constraints Majorons in the mass window $0.1 \,\text{eV}\lesssim m_\phi \lesssim 300 \,\text{eV}$. These novel constraints test coupling strengths as small as $\lambda \sim 10^{-13}$ and we note that they generically apply to any boson with a decay mode to neutrinos. 
\item A Majoron+dark radiation cosmology is capable of reducing the outstanding Hubble tension from $4.4\sigma$ to $2.5\sigma$. On the one hand, the amelioration is not perfect and some tension remains. On the other hand, the preferred region of parameter space to solve the Hubble tension is theoretically very well motivated. The region corresponds to $m_\phi \sim (0.1-1)\,\text{eV}$, $\lambda \sim (10^{-14}-10^{-13})(\text{eV}/m_\phi)$, and $\Delta N_{\rm eff} = 0.5\pm 0.2$. In this region of parameter space, majoron masses are compatible with quantum gravity expectations and $\nu\!-\!\phi$ coupling strengths correspond to scales of lepton number breaking within the type-I seesaw mechanism suggestively close to the electroweak scale, $v_L \sim v_H$. 
\end{enumerate} 

To conclude, Ref.~\cite{Escudero:2019gvw} has only partially answered the question motivating this study and that provides the title of this contribution. Looking forward, in~\cite{Escudero:2019gvw}, being maximally conservative, we restricted ourselves to an scenario in which there existed no primordial Majorons. However, Majorons could very well have been produced in the early Universe. From the perspective of the CMB, the existence of a primordial Majoron population will yield more severe constraints than those highlighted in blue in Figure~\ref{fig:Money}. However, in the context of the Hubble tension, a primordial population of Majorons can lead to a substantially enhanced expansion history of the Universe prior to recombination which could be capable of fully solving the Hubble tension. \\

\acknowledgement{This work was supported by the European Research Council under the European Union's Horizon 2020 program (ERC Grant Agreement No 648680 DARKHORIZONS).}

%%% to use a bib file  
%\bibliographystyle{JHEP}
%\bibliography{refs} 

\end{document}